\documentclass[twocolumn,floats,showpacs,amsmath,amssymb,prb]{revtex4}
\usepackage{graphicx}

\begin{document}

\title{Vibrational properties and diffusion of hydrogen on graphene}
\author{Carlos P. Herrero}
\author{Rafael Ram\'{\i}rez}
\affiliation{Instituto de Ciencia de Materiales,
         Consejo Superior de Investigaciones Cient\'{\i}ficas (CSIC),
         Campus de Cantoblanco, 28049 Madrid, Spain }
\date{\today}

\begin{abstract}
   Hydrogen and deuterium chemisorption on a single layer of graphene has
 been studied by path-integral molecular dynamics simulations.
 Finite-temperature properties of these point defects were analyzed in 
the range from 200 to 1500 K, by using a tight-binding potential
fitted to density-functional calculations. 
On one side, vibrational properties of the adatoms are studied at their
equilibrium positions, linked to C atoms. The vibrations display an
appreciable anharmonicity, as derived from comparison between kinetic
and potential energy, as well as between vibrational energy for hydrogen
and deuterium.
On the other side, adatom motion has been studied by quantum 
transition-state theory.
At room temperature, quantum effects are found to enhance the hydrogen 
diffusivity on the graphene sheet by a factor of 20.
\end{abstract}



\maketitle

\section{Introduction}

In recent years there has been a surge of interest in carbon-based
materials. Among them, those formed by C atoms with $sp^2$
hybridization have been intensively studied, as is the case of
carbon nanotubes, fullerenes, and graphene. The latter, in particular,
is up to now the only real two-dimensional crystal, with exotic
electronic properties.\cite{ge07,ka07}

Carbon-based systems, in general, are considered as candidates for
hydrogen storage.\cite{di01} Also, chemisorption on two-dimensional 
systems, such as graphene, can be important for catalytic 
processes.\cite{sl03}
The interest on hydrogen as an impurity in solids and on surfaces is not 
new, and dates back to many years.
Even though this is one of the simplest impurities, a thorough understanding 
of its physical properties is complex due to its low mass, and requires the 
combination of advanced experimental and theoretical methods.\cite{pe92,es95}

 Experimental investigations on atomic, isolated hydrogen on graphene have 
been so far scarce, since H is difficult to detect. 
Recently, it has been clearly observed by transmission electron microscopy,
and its dynamics were analyzed in real time.\cite{me08}
In general, apart from its basic interest as an isolated impurity, 
an important property of hydrogen in solids and surfaces is its ability to 
form complexes and passivate defects, which has been extensively studied 
in the last twenty years.\cite{pe92,es95,ze99} 

From a theoretical viewpoint, atomic hydrogen on graphene has been studied
by several authors using {\em ab-initio} 
methods.\cite{sl03,ya07,ca09,bo08,du04,an08}
It is generally accepted that chemisorption of a single hydrogen atom 
leads to the appearance of a defect-induced magnetic moment on the graphene 
sheet, along with a large structural distortion.\cite{ya07,ca09,bo08}
However,  standard electronic-structure calculations, in spite of 
their quantum mechanical character, usually treat atomic nuclei as classical 
particles, and typical quantum effects as zero-point vibrations are 
not directly available from the calculations.
Such quantum effects may be relevant for vibrational and electronic
properties of light impurities like hydrogen, especially at low temperatures. 

Finite-temperature properties of hydrogen-related defects in solids
have been studied by {\em ab-initio} and tight-binding (TB) molecular dynamics
simulations. In many earlier applications of these methods, atomic 
nuclei were treated as classical particles.\cite{bu91,pa94,be00} 
To consider the quantum character of the nuclei,
the path-integral molecular dynamics approach results
to be particularly suitable.  In this procedure all nuclear degrees of
freedom can be quantized in an efficient way, allowing one to include
both quantum and thermal fluctuations in many-body systems
at finite temperatures.  Thus, molecular dynamics 
sampling applied to evaluate finite-temperature path integrals allows one to 
carry out quantitative studies of anharmonic effects in 
condensed matter.\cite{gi88,ce95}

In this paper, the path-integral molecular dynamics (PIMD) method is used 
to investigate the role of
the impurity mass on the properties of hydrogenic point defects.
We study isolated hydrogen and deuterium (D) on a graphene sheet.
Special attention has been laid upon the vibrational properties of these
impurities, by considering anharmonic effects on their quantum dynamics.
 The results of these calculations show that such anharmonic effects lead
to an appreciable deviation of the vibrational energy of the 
impurities, as compared to a harmonic approximation.  
Also, the atomic diffusion is found to be enhanced with respect to
the classical limit, for both H and D.
Path-integral methods analogous to that employed in this work 
have been applied earlier to study hydrogen in metals\cite{gi88,ma95,ma95b} 
and semiconductors.\cite{ra94,he95,mi98,he07}
In connection with the present work, hydrogen has been studied inside and on 
carbon nanotubes by diffusion Monte Carlo.\cite{go01,go07}

 The paper is organized as follows. In Sec.\,II, we describe the
computational methods employed in our calculations. 
Our results are presented in Sec.\,III, dealing with the energy of the 
defects, vibrational properties, and impurity diffusion.
In Sec.\,IV we summarize the main results.

\section{Computational Methods}

In this Section we present the computational methods employed in our
simulations. On one side, in Sec.~II.A we introduce the PIMD method
used to obtain equilibrium properties related to the hydrogenic defects.
On the other side, in Sec.~II B we discuss a procedure to calculate
rate constants for impurity jumps, in the context of transition-state
theory. 

\subsection{Path-integral molecular dynamics}

Our calculations are based on the path-integral formulation of statistical 
mechanics, which is a powerful nonperturbative approach to study many-body 
quantum systems at finite temperatures. In this approach, the partition
function is evaluated through a discretization of the density matrix
along cyclic paths, composed of a number $L$ (Trotter number)
of ``imaginary-time'' steps.\cite{fe72,kl90} In the numerical simulations,
this discretization gives rise to the appearance of $L$ ``beads''
for each quantum particle.  Then, this method exploits the fact that 
the partition function of a quantum system can be written in a way
formally equivalent to that of a classical one, obtained by substituting each 
quantum particle by a ring polymer consisting of $L$ classical particles, 
connected by harmonic springs.\cite{gi88,ce95}
Here we employ the molecular dynamics technique to sample the configuration
space of the classical isomorph of our quantum system ($N$ carbon atoms
plus one impurity).
Calculations were carried out in the canonical ensemble,
using an originally developed software, which enables efficient PIMD 
simulations on parallel computers.
The algorithms employed to integrate the equations of motion were based
on those described in the literature.\cite{ma96,tu02}

The calculations have been performed within the adiabatic
(Born-Oppenheimer) approximation, which allows us to define a potential
energy surface for the nuclear coordinates.
An important question in the PIMD method is an adequate description of the 
interatomic interactions, which should be as realistic as possible.
Since employing true density functional or Hartree-Fock based
self-consistent potentials would restrict enormously the size of our
simulation cell, we have derived the Born-Oppenheimer surface for the
nuclear dynamics from an efficient tight-binding effective Hamiltonian, 
based on density functional (DF) calculations.\cite{po95}
The capability of TB methods to reproduce different properties of
molecules and solids was reviewed by Goringe {\em et al.}\cite{go97}
In particular, the reliability of our TB Hamiltonian to describe 
hydrogen-carbon interactions in carbon-based materials has been checked
in previous work,\cite{he06,he07} and according to those results we
expect that systematic errors in calculated diffusion barriers for H in 
these materials are less than 0.1 eV.
An advantage of the combination of path integrals with electronic structure 
methods is that both electrons and atomic nuclei are treated quantum 
mechanically, so that phonon-phonon and electron-phonon interactions 
are directly taken into account in the simulation.

  Simulations were carried out in the $NVT$ ensemble on a $4\times4$ 
graphene supercell of size $4 a$ = 9.84 \AA\ with periodic boundary 
conditions, containing 32 C atoms and one adatom. 
For comparison, we also carried out simulations of graphene without 
impurities, using the same supercell size.
We have checked that larger supercells, i.e., $5\times5$ give
within error bars the same results as those derived below from the
$4\times4$ supercell.
Sampling of the configuration space has been carried out 
at temperatures between 200 and 1500 K. 
 For a given temperature, a typical run consisted of $2 \times 10^4$ PIMD 
steps for system equilibration, followed by $10^6$ steps for the calculation 
of ensemble average properties. 
To have a nearly constant precision in the results
at different temperatures, we took a Trotter number that
scales as the inverse temperature, so that $L T$ = 6000 K.
For comparison with the results of our PIMD simulations, we have carried
out some classical molecular dynamics simulations with the same interatomic 
interaction, which is achieved by setting $L$ = 1.
The quantum simulations were performed using a staging transformation
for the bead coordinates.
Chains of four Nos\'e-Hoover thermostats were coupled to each degree of 
freedom to generate the canonical ensemble.\cite{tu98}
The equations of motion were integrated by using
the reversible reference system propagator algorithm (RESPA), which allows
us to define different time steps for the integration of the fast and slow
degrees of freedom.\cite{ma96} 
The time step $\Delta t$ associated to the DF-TB forces was taken in the 
range between 0.2 and 0.5 fs, which was found to be appropriate for the
atomic masses and temperatures studied here.
For the evolution of the fast dynamical variables, that include the
thermostats and harmonic bead interactions, we used a 
time step $\delta t = \Delta t/4$.
More details on the actual implementation of the simulation method
can be found elsewhere.\cite{ra06,he06}

\subsection{Quantum transition-state theory}

Classical transition-state theory (TST) is a well-established method for 
calculating  rate constants of infrequent events.
An important element in this computational method  is the ratio between 
the probability of finding the system at a barrier (saddle-point)
and at its stable configuration. 
A quantum extension of this theory has been developed in the context
of path integrals, with the purpose of studying the kinetics of processes 
involving light atoms.\cite{gi88,vo89b,ca94}
This quantum approach allows one to relate the jump rate $k$ to the
probability density of the center-of-gravity (centroid) of the 
quantum paths of the jumping atom, defined as
\begin{equation}
   \overline{\bf x} = \frac{1}{L} \sum_{i=1}^L {\bf x}_i  \, ,
\label{centr}
\end{equation}
${\bf x}_i$ being the coordinates of the ``beads'' in the associated ring
polymer.
Then, $k$ is related to the ratio $P_c$ between the equilibrium probability
of finding the centroid at a saddle-point (say ${\bf x}^*$) and at a stable
site (say ${\bf x}_0$).\cite{gi87b,vo89b}  Namely:
\begin{equation}
    k = \frac{\overline{v} P_c}{2 l} \, ,
\label{kk}
\end{equation}
where $\overline{v}$ is a factor weakly dependent on temperature,
taken to be the thermal velocity $\overline{v}=\sqrt{2 / (\pi \beta m)}$
of the jumping impurity, and $l$ is the distance between
${\bf x}_0$ and ${\bf x}^*$.
Note that, apart of typical quantum effects, the $1 / \sqrt{m}$ factor
in $\overline{v}$ will favor a faster jump rate of the lighter atoms,
as in classical TST.    The probability ratio $P_c$ 
can be written as $\text{exp}(-\beta \Delta F)$, $\Delta F$ being
an effective free-energy barrier, given by the reversible work done on the
system when the impurity centroid $\overline{{\bf x}}$ moves along a path
from ${\bf x}_0$ to ${\bf x}^*$:
\begin{equation}
 \Delta F = - \int_{{\bf x}_0}^{{\bf x}^*} {\bf f}(\overline{{\bf x}})
     d\overline{{\bf x}}  \, ,
\label{deltaf}
\end{equation}
where ${\bf f}(\overline{\bf x})$ is the mean force acting
on the impurity with its centroid fixed on $\overline{{\bf x}}$
at temperature $T$:
\begin{equation}
{\bf f}(\overline{{\bf x}}) = - \langle \nabla_{\bf x} V({\bf R})
                  \rangle_{\overline{\bf x}}  \, .
\label{fx}
\end{equation}
Here $V({\bf R})$ is the potential energy, ${\bf R}$ being in our case a
$3(N+1)$-dimensional vector ($N$ carbon atoms plus one impurity).
The average value in Eq.~(\ref{fx}) is taken over quantum paths
with the centroid of the impurity fixed on $\overline{{\bf x}}$
at temperature $T$.
Thus, the jump rate (a dynamical quantity) is related to the free energy 
difference $\Delta F$ (a time-independent quantity), so that its temperature 
dependence can be obtained from equilibrium simulations without any
direct dynamical information.\cite{vo93}
Quantum effects that may give rise to substantial deviations
from the classical jump rate are taken into account
within this kind of quantum TST.
In this way, jump rates for kinetic processes can be obtained for 
realistic, highly nonlinear many-body problems. 
The reliability of this method to calculate free-energy barriers and
jump rates has been discussed in Refs. \onlinecite{gi87b,ma95}.
In particular, it was argued that this method can be inaccurate in
the presence of asymmetric barriers at low temperatures. In our case,
the error bar of the calculated barriers is expected to be less than
the intrinsic error of the employed tight-binding potential.

In this kind of simulations, adatom motion was restricted along
a reaction coordinate defined by the centroid of the quantum path.
Note, for comparison, that in the equilibrium PIMD simulations
described in the previous section, there is no restriction on the
motion of the H (D) centroid.
 The force ${\bf f}(\overline{{\bf x}})$ has been 
evaluated at 11 points along the reaction coordinate  connecting the 
lowest-energy configuration (H linked to
a C atom) with the saddle point of the energy surface, and then the 
integral in Eq.~(\ref{deltaf}) was calculated numerically.
For each point in the integration path, we generated 5000 configurations
for system equilibration, and 2 $\times$ $10^4$ configurations for 
calculating the mean force at a given temperature. More technical details
can be found elsewhere.\cite{gi88,he97,no97b}

\section{Results}

\subsection{Vibrational energy}

We first discuss the lowest-energy configuration for the hydrogenic 
impurities on a graphene sheet, as derived from classical calculations 
at $T = 0$, i.e., point atomic nuclei without spatial delocalization. 
The impurity binds to an C atom, which relaxes out of the sheet plane
by 0.46 \AA, with a bond distance between C and impurity of 1.17 \AA.
These results are in line with those reported in the literature,
and in particular with the breaking of a $\pi$ bond and producing an
additional $\sigma$ bond, changing the hybridization of the involved
C atom from $s p^2$ to $s p^3$.\cite{sl03,bo08,ca09}
Assuming the host C atoms fixed in the relaxed geometry,
one can calculate vibrational frequencies for the impurity in a
harmonic approximation. Thus, we find for hydrogen a frequency
$\omega_{\perp}$ = 2555 cm$^{-1}$ for stretching of the C--H
bond (perpendicular to the graphene sheet), and 
$\omega_{\|}$ = 1186 cm$^{-1}$ for vibrations parallel to the
plane (twofold degenerate). 

We now turn to the results of our simulations at finite temperatures,
and will discuss the energy of the hydrogenic defects..
The internal energy of the system graphene plus impurity, $E(T)$, at 
temperature $T$ can be written as:
\begin{equation}
  E(T) =  E_{\rm min} + E_{\rm v}(T)   \, ,
\label{et}
\end{equation}
where $E_{\rm min}$ is the potential energy for the classical material
at $T = 0$ (point-like atoms on their equilibrium positions), and 
$E_{\rm v}(T)$ is the vibrational energy of the whole system.
Then, $E_{\rm v}(T)$ can be obtained by subtracting the energy 
$E_{\rm min}$ from the internal energy derived from PIMD simulations.
 In Fig.~1 we show the temperature dependence of the vibrational 
energy $E_{\rm v}$ for a $4 \times 4$ graphene supercell including 
an H atom (circles).
For comparison we also show $E_{\rm v}$ for a pure graphene sheet (squares).  
At 300 K, the vibrational energy of graphene
amounts to 6.18 eV per simulation cell, i.e., 0.19 eV/atom. 
As expected, $E_{\rm v}$ increases as temperature is raised, and eventually
converges to the classical limit $E_{\rm v}^{cl} = 3 N k_B T$ at high $T$.

\begin{figure}
\vspace{-2.0cm}
\hspace{-0.5cm}
\includegraphics[width= 9cm]{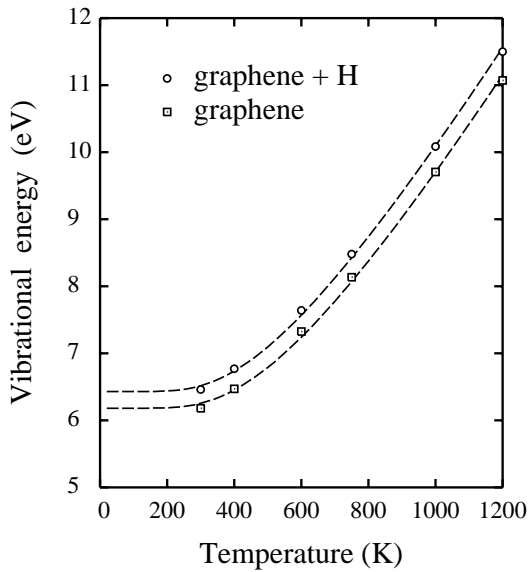}
\vspace{-2.5cm}
\caption{
Temperature dependence of the vibrational energy $E_{\rm v}$ of the
$4\times4$
graphene supercell with one hydrogen, as derived from PIMD simulations
(circles).  For comparison, we also present results for a pure graphene
supercell (squares).  Dashed lines are guides to the eye.
Error bars are in the order of the symbol size.
}
\label{f1}
\end{figure}

An interesting characteristic of the different hydrogenic defects (H or D)
is their associated vibrational energy.  At a given temperature, this
energy is defined as the difference 
$\Delta E_{\rm v}$ = $E_{\rm v}$(32C + Imp) -- $E_{\rm v}$(32C), 
where `Imp' stands for H or D.
Note that $\Delta E_{\rm v}$ defined in this way is not just the 
vibrational energy of a given adatom on graphene, but it also includes 
changes in the vibrations of the nearby C atoms.   
Shown in Fig.~2 is the vibrational energy associated to the 
hydrogenic adatoms as a function of temperature. Symbols indicate
results of PIMD simulations for H (squares) and D (circles), and
dashed lines are guides to the eye. For comparison, we also present
as solid lines the dependence of the vibrational energy of a single
particle in a harmonic approximation with the frequencies $\omega_{\|}$ 
and $\omega_{\perp}$ given above.
At low $T$ the actual zero-point vibrational energy results to be 
smaller than that given in the harmonic approach
[i.e., $\hbar (\omega_{\|} + \omega_{\perp}/2)$], but this trend changes 
as temperature rises. In fact, at temperatures larger than 1000 K the 
vibrational energy derived from the simulations is larger than that
obtained for the one-particle harmonic approach.  

\begin{figure}
\vspace{-2.0cm}
\hspace{-0.5cm}
\includegraphics[width= 9cm]{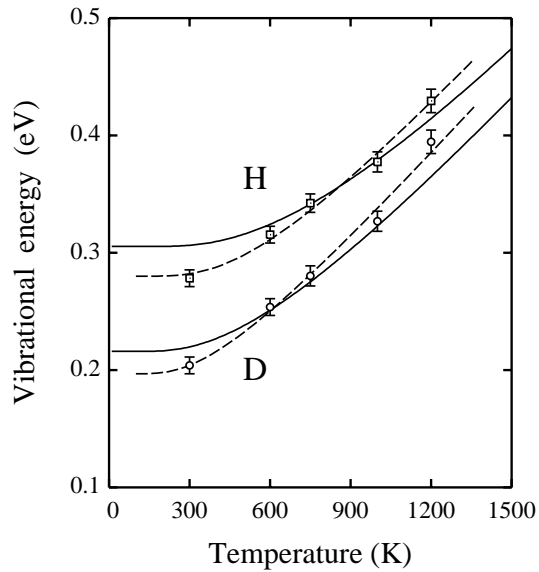}
\vspace{-2.5cm}
\caption{
 Vibrational energy $\Delta E_{\rm v}$  of the hydrogenic defects as a
function of temperature.
Results are shown for hydrogen (squares) and deuterium (circles).
 For comparison, we also present the vibrational energy obtained in a
harmonic approximation with frequencies $\omega_{\|}$ and $\omega_{\perp}$
corresponding to H and D (solid lines).
Dashed lines are guides to the eye.
}
\label{f2}
\end{figure}

Another way of getting insight into the anharmonicity of the defect
vibrations is by looking at the ratio 
$\Delta E_{\rm v}^{\text H} / \Delta E_{\rm v}^{\text D}$
between vibrational energies of point defects for both isotopes. 
At low temperature, this ratio should converge to $\sqrt{2}$ in the 
harmonic approximation, going to unity at high temperatures.
The results yielded by our simulations are shown in Fig.~3,
along with the one-particle harmonic expectancy.
The former lie somewhat lower than the latter in the whole temperature
region under consideration.
In particular, at 300 K we find 
$\Delta E_{\rm v}^{\text H} / \Delta E_{\rm v}^{\text D}$ = 1.364, 
vs a ratio of 1.394 derived in a harmonic approximation.

\begin{figure}
\vspace{-2.0cm}
\hspace{-0.5cm}
\includegraphics[width= 9cm]{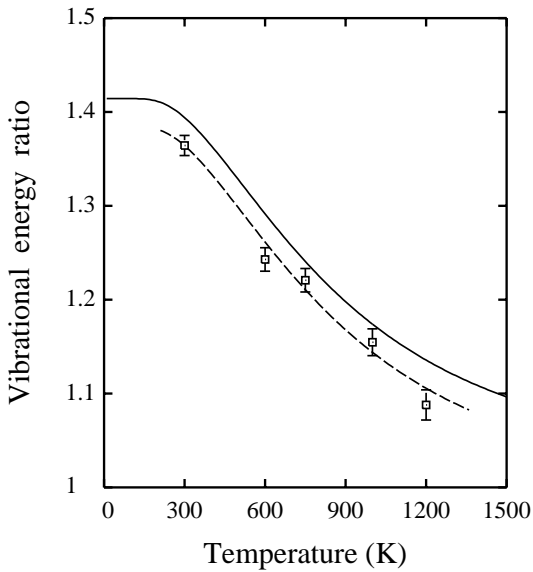}
\vspace{-2.5cm}
\caption{
Temperature dependence of the ratio
$\Delta E_{\rm v}^{\text H} / \Delta E_{\rm v}^{\text D}$
between vibrational energies of hydrogen and deuterium  defects.
Symbols indicate results of PIMD simulations.
For comparison, we also present the ratio expected in a one-particle
harmonic approximation (continuous line). The dashed line is a guide
to the eye.
}
\label{f3}
\end{figure}

PIMD simulations  allow us to separate the potential ($E_p$) and kinetic
($E_k$) contributions to the vibrational energy.\cite{he82,gi88b,gi90}
In fact, $E_k$ is related to the spatial delocalization of the quantum
paths, which can be obtained directly from the simulations (see below).
 In Fig.~4, we present the kinetic and potential energy of the 
point defect associated to hydrogen, as a function of temperature.
Symbols indicate results of our simulations for $E_k$ (squares)
and $E_p$ (circles), whereas dashed lines are guides to the eye.
The potential energy of the point defect is found to be clearly larger
than the kinetic energy, indicating an appreciable anharmonicity of the
whole defect (in a harmonic approach one has $E_k = E_p$). A solid line 
represents the expected dependence for both,
kinetic and potential contributions, in a harmonic approximation with
the frequencies $\omega_{\|}$ and $\omega_{\perp}$ given above.
At low temperature we find an appreciable change of the kinetic energy
respect to the harmonic approach, contrary to the potential energy, which
coincides within error bars with the harmonic expectancy. 
A qualitative understanding of this behavior can be obtained by 
analyzing  the energy changes obtained through time-independent
perturbation methods.\cite{la65,he95}  Thus, assuming a perturbed 
one-dimensional harmonic oscillator (with perturbations
of $x^3$ and $x^4$ type) at $T = 0$, the first-order change in the 
energy is totally due to a change in the kinetic energy, the potential
energy remaining unaltered respect to its unperturbed value. 
A similar behavior has been obtained for hydrogen in silicon from
path-integral Monte Carlo simulations.\cite{he95} The main difference is 
that in that case the kinetic energy was found to increase respect to the
harmonic value, contrary to the result obtained here for H on graphene.
This seems to depend on the details of the interatomic interactions and
the actual geometry of the point defect under consideration, but in both
cases the potential energy at low temperature is very close to the value 
yielded by the harmonic approximation.

\begin{figure}
\vspace{-2.0cm}
\hspace{-0.5cm}
\includegraphics[width= 9cm]{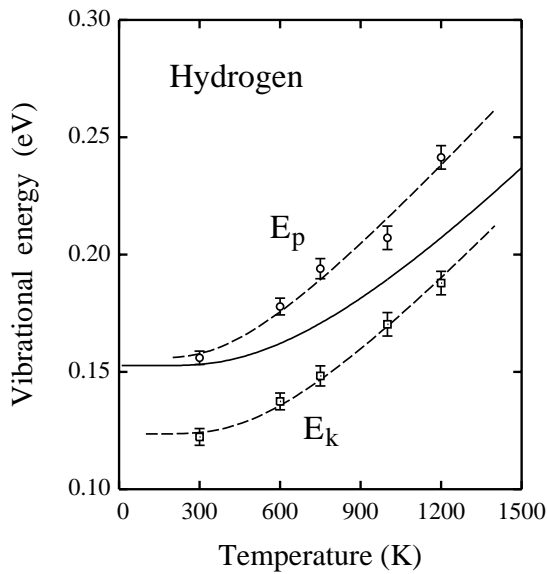}
\vspace{-2.5cm}
\caption{
Temperature dependence of kinetic (squares) and potential (circles)
contributions to the vibrational energy of the H defect.
A solid line represents the expectancy of a harmonic approximation
with frequencies $\omega_{\|}$ and $\omega_{\perp}$.
 Dashed lines are guides to the eye.
}
\label{f4}
\end{figure}

Path-integral simulations at finite temperatures describe 
quantum delocalization through paths of finite size. This means that
the average extension of the paths is a measure of the importance of
quantum effects in a given problem. One can define a kind of
quantum delocalization as the mean-square ``radius of gyration''
$D^2$ of the ring polymers associated to the quantum particle under
consideration.\cite{gi88,gi90} This means:
\begin{equation}
D^2 = \frac{1}{L} \left< \sum_{i=1}^L ({\bf x}_i - \overline{\bf x})^2
           \right>    \, ,
\end{equation}
where $\overline{\bf x}$ is the position of the centroid
of the paths defined in Eq.~(\ref{centr}), 
and $\langle ... \rangle$ indicates a thermal average at temperature $T$.
Note that the total spacial delocalization of a particle includes an
additional term, taking into account displacements of the center of gravity
of the paths. This term is the only one surviving at high temperatures,
since in the classical limit each path collapses onto a single point.
For our problem of hydrogen on graphene, with well-defined H vibrations
along three orthogonal axes, one can define a quantum delocalization of
hydrogen along
each of these directions ($D^2_{\|}$ and $D^2_{\perp}$).
These delocalizations are displayed in Fig.~5 for vibrations 
parallel (circles) and perpendicular (squares) to the graphene plane, 
as derived  from our PIMD simulations.
$D^2$ decreases as the temperature is raised and the particle becomes
more ``classical''.
For comparison, we also show  in Fig.~5  the mean-square displacement 
$D^2$ expected for harmonic 
oscillators of frequencies $\omega_{\|}$ and $\omega_{\perp}$, which can be 
worked out analytically.\cite{gi88,gi90,ra93}
The delocalization $D^2_{\|}$ derived from the simulations follows
closely the harmonic expectancy in the whole temperature range considered
here. However, $D^2_{\perp}$ is higher than its corresponding harmonic
result at $T <$ 400 K. 

\begin{figure}
\vspace{-2.0cm}
\hspace{-0.5cm}
\includegraphics[width= 9cm]{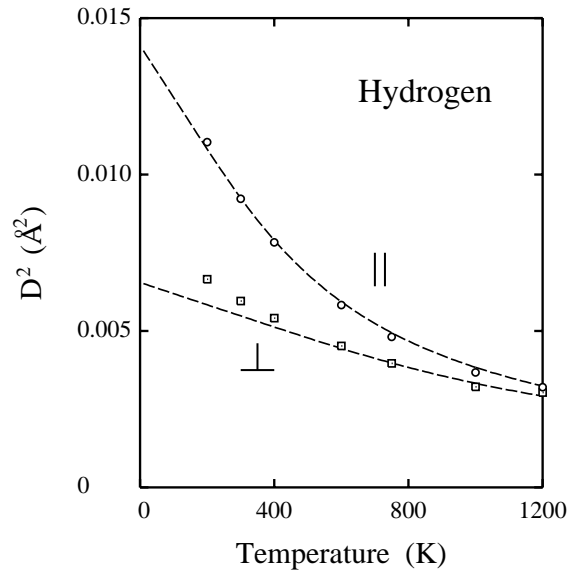}
\vspace{-2.5cm}
\caption{
Mean-square displacement $D^2$ of the quantum paths for hydrogen in
directions parallel (circles) and perpendicular (squares) to the
graphene sheet.
 Dashed lines correspond to a harmonic approximation for vibrations
with frequencies $\omega_{\|}$ and $\omega_{\perp}$.
Error bars of the simulations results are on the order of the symbol
size.
}
\label{f5}
\end{figure}

\subsection{Hydrogen diffusion} 

Hydrogen is expected to diffuse on the graphene sheet, breaking a
C--H bond and forming a new one with a nearby C atom.
To check the appearance of this kind of hydrogen jumps, we have carried
out some classical molecular dynamics simulations, which allow one
to follow the adatom along its trajectory on the surface. As indicated
above, this kind of classical limit is easily achieved with our PIMD
code, by setting the Trotter number $L =1$ (one bead per atom).
At temperatures lower than 1000 K, hydrogen jumps on the graphene surface
result to be infrequent events, and are rarely observed 
along a simulation run. Then, a reliable estimation of the diffusion
coefficient by this method is not possible.
Moreover, at temperatures larger than 1000 K hydrogen begins to 
escape from the sheet, without remaining on it enough time 
for studying quantitatively the diffusion process. 
This situation is remedied by analyzing the hydrogen
motion by TST, and in particular by the quantum version presented in
Sect.~II.B. This procedure allows us to obtain free energy barriers
for impurity jumping, which include corrections due to the quantum
character of the atomic nuclei, and in particular the renormalization
of the barriers caused by zero-point motion. Also, phonon-assisted
tunneling is included in the calculation, since motion of the C
atoms takes into account a full quantization of the vibrational 
degrees of freedom.  
The main question we address here is the dependence of hopping rates on
temperature and impurity mass. In connection with this, there is a
vast literature about theoretical models for quantum diffusion of light
particles in solids.\cite{fl70,su80,sc88,gi88}
Due to the complexity of this problem, such calculations have been 
typically based on model potentials for the impurity-lattice interactions.

Here, the jump rate of the impurity between two nearest equilibrium sites is
derived from the free-energy barrier between those sites.
To calculate this barrier, we first select a continuous path from one 
site to the other, minimizing the energy at the transition (saddle)
point in a classical calculation (point-like atoms) at $T = 0$.
In connection with this, it is worthwhile noting that seemingly
simple atomic jumps can actually involve coupled barriers, as 
indicated in Ref.~\onlinecite{ra96}. Thus, to obtain the barrier for
hydrogen jumps, we considered a coupled motion of H and the nearest C 
atom. For the optimal path, we obtained an energy barrier of 0.78 eV,
with the saddle point corresponding to a symmetric (bridge) configuration 
of hydrogen between two C atoms and at a distance of 1.7 \AA\ from the 
graphene plane.
Once selected the best path, we performed finite-temperature 
 path-integral simulations with the centroid of H fixed on several
points along this path, as described in Sect.~II.B. Then, the free-energy 
barrier is calculated from the mean force by using Eq.~(\ref{deltaf}). 

In Fig.~6 we present the free-energy barrier $\Delta F$
for adatom diffusion between adjacent C--H bonds.
Data derived from line integration of the mean force
are shown as a function of temperature for hydrogen (squares) and
deuterium (circles). For comparison, we also show results derived from
classical simulations (triangles).
In this plot, one notices first that $\Delta F$
is higher for D than for H, and even higher for the classical limit,
which in this respect may be considered as the large-mass limit
$M \to \infty$.\cite{ra06,he06}
Then, we observe an appreciable decrease in the free-energy barrier
as the impurity mass is reduced. 
Second, one observes an increase in $\Delta F$ as temperature is
raised in the three cases shown in Fig.~6. This increase is smaller,
but not negligible, in the classical limit, which at $T \to 0$
converges to the potential energy barrier given above 
($\Delta E_p$ = 0.78 eV).  
At low temperature, the dependence of $\Delta F$ upon impurity mass
is related to the change in internal energy $E$ of the
defect complex along the diffusion path, since then the entropy 
contribution to the free energy becomes negligible.
At 300 K we find $\Delta F$ = 0.79, 0.74, and 0.71 eV, for the 
classical limit, D, and H, respectively. This means that quantum effects
renormalize the free-energy barrier at room temperature by about 6~\% 
for D and 10~\% for H.
This reduction in $\Delta F$ decreases as temperature rises and the
atoms become ``more classical''. In fact, these free-energy barriers 
should converge one to the other in the high-temperature limit, 
which cannot be approached here due to the onset of adatom desorption.

\begin{figure}
\vspace{-2.0cm}
\hspace{-0.5cm}
\includegraphics[width= 9cm]{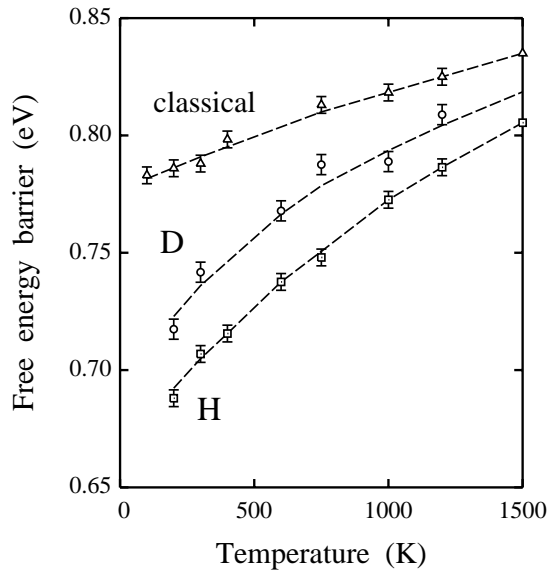}
\vspace{-2.5cm}
\caption{
Effective free-energy barrier for impurity jumps on the graphene
sheet as a function of temperature. Squares, hydrogen; circles, deuterium;
triangles, classical limit.  Dashed lines are guides to the eye.
}
\label{f6}
\end{figure}

Shown in Fig.~7 is the rate for impurity jumps on a graphene
sheet as a function of the inverse temperature. Data were derived from
the free-energy barriers displayed in Fig.~6, by using Eq.~(\ref{kk}).
Results are given for hydrogen (squares), deuterium (circles), and
classical limit (triangles).
At $T \le 500$ K,
the jump rate for hydrogen results to be larger than that for deuterium,
which in turn is higher than that found in the classical limit, as
expected from the change in effective free-energy barrier discussed
above.
At 300 K, the jump rate for hydrogen is found to be 11.6 s$^{-1}$, about
20 times larger than the value found in the classical calculation.
The influence of quantum effects on hydrogen diffusivity increases as
temperature is lowered, as could be expected, but the jump rate itself
becomes very small. In fact, at 200 K the calculated rate for hydrogen,
$k_{\text H}$, is less than $10^{-4}$ s$^{-1}$. 
On the contrary, at high temperatures $k_{\text H}$  converges
to the classical result, and the difference between both becomes
unobservable at $T$ larger than 1000 K. At this temperature we
find $k_{\text H} = 2.0 \times 10^9$ s$^{-1}$.

\begin{figure}
\vspace{-2.0cm}
\hspace{-0.5cm}
\includegraphics[width= 9cm]{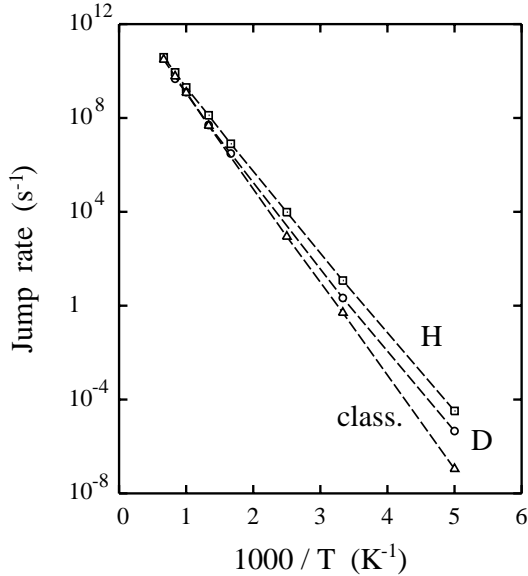}
\vspace{-2.5cm}
\caption{
Rate for impurity jumps on the graphene sheet as a function of the
inverse temperature. Symbols represent results derived from
simulations for hydrogen (squares), deuterium (circles), and
classical limit (triangles).
 Dashed lines are guides to the eye.
}
\label{f7}
\end{figure}

The diffusion barriers obtained here for H and D on graphene are
comparable to those found for hydrogen diffusion on graphite.
Ferro {\em et al.}\cite{fe04} used density-functional theory to
study various diffusion barriers in graphite.
For hydrogen diffusion on its surface, they obtained  a (classical) 
barrier of 0.94 eV, somewhat larger than that found here at $T = 0$ in
the classical limit.

The diffusion of H and D on solid surfaces has been studied
earlier using quantum TST, employing simulations similar to those
presented here. Such studies were mainly focused on hydrogen
diffusion on metal surfaces.\cite{ma95b,ri93} Our results are
qualitatively similar to those obtained in these studies, with
an appreciable enhancement of the jump rate of the impurity in
comparison to a classical model.
At temperatures lower than those studied here ($T <$ 50~K),
a temperature-independent diffusion was found on metal 
surfaces.\cite{ma95b,ma97} This is also a possibility for hydrogen
diffusion on graphene, and remains as a challenge for future 
research. PIMD simulations at those temperatures would need the
use of Trotter numbers $L$ larger than 100, that together with the
interaction potentials employed here requires computational 
resources out of the scope of the present work.

\section{Summary}

The main advantage of PIMD simulations of hydrogen on graphene 
is the possibility of calculating defect energies at finite temperatures,
including a full quantization of host-atom motions, which are not 
easy to take into account from fixed-lattice calculations and
classical simulations. 
Isotope effects can be readily explored, since the impurity mass
appears as an input parameter in the calculations. This includes the
consideration of zero-point motion, which together with anharmonicity
may cause appreciable non-trivial effects.
 Our results indicate that hydrogen adsorbed on graphene
cannot be accurately described as a particle moving in a harmonic potential. 
Even if anharmonicities of the interatomic potential are taken into account,
a single-particle approximation is questionable as a realistic description
of impurity complexes at finite temperatures. It is then necessary to treat
the defect as a many-body problem with anharmonic interactions. 
                                                                                    
{\em Ab-initio} theoretical techniques to calculate defect energies in 
solids have achieved an excellent precision in recent years. 
 However, zero-point motion is a factor limiting the accuracy of 
state-of-the-art techniques to predict energy bands and total energies
of solids.\cite{ra06}
The same happens for defect levels caused by impurities in solids, 
since their energy may change appreciably as the impurity mass is varied. 
Here we have illustrated how anharmonicities in the atomic motion
cause an appreciable difference between kinetic and potential energy 
of the defect (about 20 \% for H at 300 K), and have quantified the 
effect of the impurity mass on anharmonic shifts in the energy.

Due to the large relaxation of the nearest C atoms,
hydrogen migration requires important motion of these atoms. 
Then, an adatom jump has to be viewed as a cooperative process
involving a coupled motion of the impurity and the nearest host atoms.  
This picture is similar to that described in the literature as 
``opening of a door'',\cite{bo94} which favors impurity diffusion.
Thus, quantum motion of both adatom and C atoms helps to renormalize
the diffusion barriers respect to the classical expectancy.
At 300 K we find a hydrogen diffusivity 20 times larger than that
derived from classical barriers.

\begin{acknowledgments}
This work was supported by Ministerio de Ciencia e Innovaci\'on (Spain) 
through Grant No. BFM2003-03372-C03-03. 
\end{acknowledgments}

\end{document}